\documentclass[aps,prl,twocolumn,preprintnumbers,amsmath,amssymb,superscriptaddress]{revtex4}

\usepackage{graphicx}

\usepackage{pifont} 
\usepackage{wasysym} 

\begin{document}

\title{Minimal resonator loss for circuit quantum electrodynamics}

\author{R. Barends}
\author{N. Vercruyssen}
\author{A. Endo}
\affiliation{Kavli Institute of NanoScience, Faculty of Applied
Sciences, Delft University of Technology, Lorentzweg 1, 2628 CJ
Delft, The Netherlands}

\author{P. J. de Visser}
\affiliation{Kavli Institute of NanoScience, Faculty of Applied
Sciences, Delft University of Technology, Lorentzweg 1, 2628 CJ
Delft, The Netherlands}

\affiliation{SRON Netherlands Institute for Space Research,
Sorbonnelaan 2, 3584 CA Utrecht, The Netherlands}

\author{T. Zijlstra}
\author{T. M. Klapwijk}
\affiliation{Kavli Institute of NanoScience, Faculty of Applied
Sciences, Delft University of Technology, Lorentzweg 1, 2628 CJ
Delft, The Netherlands}

\author{P. Diener}
\author{S. J. C. Yates}
\author{J. J. A. Baselmans}
\affiliation{SRON Netherlands Institute for Space Research,
Sorbonnelaan 2, 3584 CA Utrecht, The Netherlands}

\date{\today}

\begin{abstract}
Single photon level quality factors of $500\cdot 10^3$ are shown in
NbTiN superconducting resonators at millikelvin temperatures. This
result originates from the intrinsic low dielectric loss of NbTiN,
as demonstrated by comparison with Ta, and by removing unnecessary
parts of the dielectric substrate.
\end{abstract}

\maketitle

In circuit quantum electrodynamics quantum information processing is
done by coupling the qubit state to a single photon bound to a
superconducting resonator \cite{wallraff}. The lifetime of a single
photon with frequency $f$ is defined by: $\tau=Q/2\pi f$
\cite{wang}, therefore a high resonator quality factor ($Q$) is
needed to maximize the lifetime. Presently used resonators, made
from Nb or Al, have quality factors on the order of $10^4$ to $10^5$
\cite{wang,oconnell,palacios,macha,lindstrom}. In contrast,
superconducting resonators for astronomical photon detection
\cite{day} have shown quality factors in excess of a million.
However, these quality factors are measured in the many-photon
regime. One would like to maintain these high values down to the
single photon level. Therefore, we study the unloaded quality factor
of NbTiN and, for comparison, Ta quarterwave resonators down to the
single photon level. NbTiN has been shown to follow Mattis-Bardeen
theory more closely than Nb, Al or Ta, indicating it has a minimal
dielectric layer compared to the latter materials \cite{barendsapl}.
We find that in the single photon regime the quality factor of NbTiN
resonators is so high that the loss is largely due to the exposed
substrate surface. In contrast, for Ta resonators the quality factor
is limited by the metal surface. We show that a further reduction of
the loss in NbTiN resonators is achieved by removing the substrate
from the regions with a high electric field density. This increases
the quality factor to half a million for resonators with a central
line width of 6 $\mu$m, three times higher than recently reported
for Re \cite{wang}.

\begin{figure}[!b]
    \centering
    \includegraphics[width=1\linewidth]{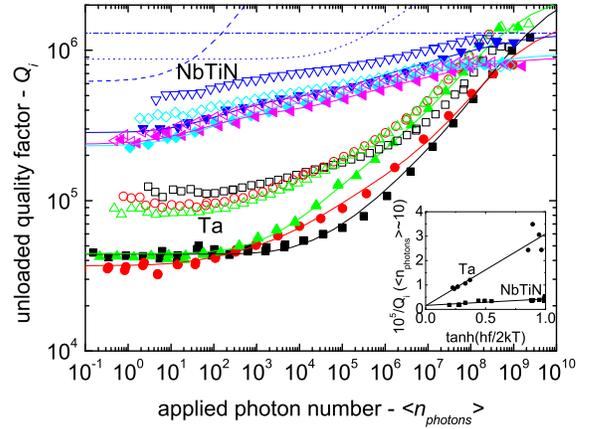}
    \caption{(Color online) The unloaded quality factor of NbTiN and Ta quarterwave resonators versus applied microwave photon number in the resonator.
    Bath temperatures are 60 mK (closed symbols) and 310 mK (open
    symbols). Central line width is $S=3$ $\mu$m and gap width is $W=2$
    $\mu$m. Frequencies of the resonators used are 3.7 ($\blacktriangledown$), 4.2 (\ding{117}) and 6.2 ($\blacktriangleleft$) GHz for the NbTiN,
    and 3.2 ($\blacksquare$), 4.5 (\ding{108}) and 5.0 ($\blacktriangle$) GHz for Ta.
    The solid lines are fits using Eq. \ref{eq:qtls}.
    The quality factors of the metal surfaces (dashed),
    exposed substrate surface (dotted) and a fixed loss term (dash-dotted) are shown for the 3.7 GHz NbTiN data.
    The inset shows the microwave loss in the single photon regime versus $\tanh(hf/2kT)$.}
    \label{fig:qstandard}
\end{figure}

We use NbTiN quarterwave coplanar waveguide resonators which are
capacitively coupled to a feedline \cite{day,barendsapl}. This
allows extracting the unloaded quality factor from the feedline
transmission. For comparison we have also made Ta resonators. The
NbTiN films, 300 and 50 nm thick, are DC sputter deposited on a
hydrogen passivated high resistivity ($>1$ k$\Omega$cm)
$\left<100\right>$-oriented Si wafer. The NbTi target used contains
70 at. \% Nb and 30 at. \% Ti. Patterning is done by reactive ion
etching in an SF$_6$/O$_2$ plasma. For the 300 nm thick film the
critical temperature is $T_c=14.7$ K, the low temperature
resistivity is $\rho=161$ $\mu\Omega$cm and residual resistance
ratio $RRR = 0.94$. For the 50 nm thick film: $T_c=13.6$ K,
$\rho=142$ $\mu\Omega$cm and $RRR = 0.96$. The 150 nm thick Ta film
($T_c=4.43$~K, $\rho=8.4$ $\mu\Omega$cm and $RRR = 3.0$) is
sputtered on a similar wafer and patterned in a CF$_4$/O$_2$ plasma.
The devices are cooled to 310 mK using a He-3 sorption cooler, with
the sample space magnetically shielded by a superconducting shield,
and down to 60 mK using an adiabatic demagnetization refrigerator
with the sample space shielded by an outer cryoperm and inner
superconducting shield \cite{barendsthesis}. Measurements have been
done using a vector network analyzer which is locked to a frequency
standard. A microwave isolator is placed in front of the low noise
amplifier.

The unloaded quality factor of NbTiN and Ta resonators is plotted
versus applied photon number \cite{numberofphotons} in the resonator
in Fig. \ref{fig:qstandard}. The resonators have resonance
frequencies in the 3-6 GHz range, a central line width of $S=3$
$\mu$m and a gap width of $W=2$ $\mu$m. Bath temperatures are 60~mK
and 310~mK. In the many-photon regime, quality factors between
$0.8\cdot 10^6$ and $1.5\cdot 10^6$ are observed for both materials.
In addition, in this regime the 60~mK and 310~mK data overlap. With
decreasing applied photon number the quality factors decrease. For
NbTiN resonators, a weak intensity dependence is observed and
quality factors decrease to $\sim 250\cdot 10^3$ at 60~mK in the
single photon regime. On the other hand, Ta quality factors degrade
quickly, decreasing to $\sim 40\cdot 10^3$. Interestingly, an
inflection point is visible in the NbTiN data around
$\left<n_{photons}\right>=10^2 - 10^3$, whereas Ta data show a
plateau at low intensities. Additionally, at low intensities a
temperature and frequency dependence develops for both materials. At
310 mK (open symbols) the quality factors are increased, for
resonators with lower frequencies the increase is larger.

Previously, we have shown that NbTiN resonators contain fewer dipole
two-level systems (TLS) than Ta, by measurements of the resonator
frequency temperature dependence \cite{barendsapl}. Dipole TLS are
configurational defects with dipole moment $p$ which reside in
amorphous dielectrics \cite{phillips}, such as native oxides.
Dielectric loss at low temperatures ($kT<hf$) arises from resonant
absorption: $1/Q \propto \tanh(hf/2kT) / \sqrt{1+E^2/E_s^2}$
\cite{phillips,martinis}. The factor $\tanh(hf/2kT)$ reflects the
thermal population difference between the lower and upper level.
With increasing intensity TLS are excited, lowering the loss. The
saturation field $E_s=\hbar/p\sqrt{T_1 T_2}$ is controlled by the
dipole moment and relaxation times $T_1$ and $T_2$.

The microwave loss of our resonators in the single photon regime
scales with $\tanh(hf/2kT)$, see the inset of Fig.
\ref{fig:qstandard}, consistent with resonant absorption from TLS.
Moreover, different resonators made from the same material follow
the same trend, indicating that the loss is very comparable over the
whole chip. In addition, the slope for Ta is steeper than for NbTiN,
consistent with a larger TLS density for Ta, compared to NbTiN
resonators.

In order to identify the location of these TLS and quantify the
influence of their saturation on the quality factor, we calculate
the effect of a hypothetical thin dielectric layer with thickness $t
\rightarrow 0$ containing TLS. Dielectric loss in a quarterwave
resonator due to a distribution of dipole TLS is given by
\cite{wang,gao}
\begin{align}
\label{eq:qtls}
\frac{1}{Q_{TLS}} = \frac{\tanh \left(\frac{hf}{2kT}\right)}{Q_{TLS,0}} \frac{ \frac{1}{2} \epsilon_0 \epsilon_{h} \iiint_{V_{h}}
\frac{|\vec{E}(\vec{r})|^2}{\sqrt{1+|\vec{E}(\vec{r})|^2/E_s^2} }
d\vec{r}} {\frac{1}{4} C V_r^2 l}
\end{align}
with $V_r$ the standing wave voltage inside the resonator, $l$ its
length and $C$ the capacitance per unit length. The dielectric loss
of the layer is $1/Q_{TLS,0}=N \pi p^2/3\epsilon_0\epsilon_h$, with
$N$ the TLS density of states and $V_h$ and $\epsilon_h$ the volume
and relative permittivity of the dielectric layer hosting the TLS.

\begin{figure}[!t]
    \centering
    \includegraphics[width=1\linewidth]{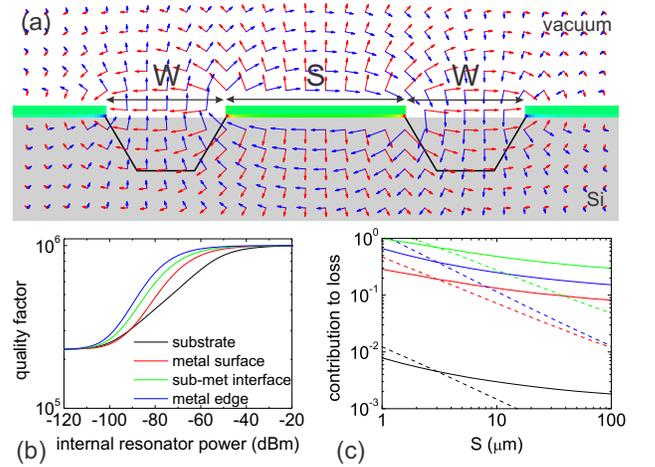}
    \caption{(Color online) (a) the charge distribution (red denotes a positive charge, blue a negative and green a neutral one),
    electric fields (red arrows) and magnetic fields (blue arrows) in the coplanar waveguide geometry.
    (b) The power dependence using Eq. \ref{eq:qtls} for a TLS distribution placed on the
    exposed substrate surface, top metal surface, substrate-metal (sub-met) interface and etched metal edges.
    $Q_0=10^6$, and at low intensity each surface is assumed to limit the $Q$ to $300\cdot 10^3$.
    (c) The normalized contribution to loss of the dielectric layers versus central line width $S$,
    for $W=2$ $\mu$m (solid) and $W=\frac{2}{3}S$ (dashed).}
    \label{fig:cpw}
\end{figure}

The electric fields for our resonator geometry are calculated by
using the potential matrix $\mathbf{P}$ to find the charge density
$q$: $\mathbf{V} = \mathbf{P} \mathbf{q}$
\cite{wang,finiteelements}. The substrate is included using the
method of partial image charges. The potential matrix elements are
given by $P_{ij} = P_{ji} = - \left(\ln|r_i - r_j| + K
\ln|r_i^*-r_j|\right) / 2\pi\epsilon_0$ for $i\neq j$, and $P_{ii} =
-\left(\ln a + K \ln[|r_i - r_i^*|+a] \right) / 2\pi\epsilon_0$,
with $r_i$ the location of the $i$-th element, $r_i^*$ the location
of the $i$-th element mirrored in the plane of the substrate
surface, $a$ its radius, $K=(1-\epsilon_s)/(1+\epsilon_s)$ and
$\epsilon_s$ the relative permittivity of the substrate. The
electric fields and magnetic fields are shown in Fig.
\ref{fig:cpw}a.

We place this hypothetical layer on either the exposed substrate
surface, top metal surface, etched metal edges or at the
substrate-metal interface. Interestingly, when the dielectric layer
is placed on any of the metal surfaces, its contribution to the loss
is two orders of magnitude larger than when placed on the exposed
substrate surface (Fig. \ref{fig:cpw}c). This can be attributed to
the high electric fields near the metal surfaces. In addition, a
dielectric placed in the vicinity of the metal leads to a much
stronger power dependence than when the dielectric layer is located
on the exposed substrate surface (Fig. \ref{fig:cpw}b). We make use
of this to distinguish between surfaces. Furthermore, the quality
factor increases with increasing central line width, irrespective of
the location of the dielectric.

In Fig. \ref{fig:qstandard}, we show that the power dependence of
the quality factor arises from the superposition of loss (solid
lines) from TLS (Eq. \ref{eq:qtls}) located at the metal surfaces
(dashed line) as well as at the exposed substrate surface (dotted
line). Interestingly, for NbTiN resonators the exposed substrate,
together with the metal surfaces, is a significant contributor to
the microwave loss. This superposition of loss closely describes the
observed point of inflection at $\left<n_{photons}\right>=10^2 -
10^3$ as well.

\begin{table}[b!]
\centering \caption{The quality factor of the dielectric layer
containing TLS, its saturation field and the additional loss factor
for the superconducting metals and for their Si substrates, used for
fitting the data in Fig. \ref{fig:qstandard}, using Eq.
\ref{eq:qtls} and $1/Q = 1/Q_0 +
1/Q_{TLS,\mathrm{met}}(E_{s,\mathrm{met}}) +
1/Q_{TLS,\mathrm{sub}}(E_{s,\mathrm{sub}})$. Calculations have been
done for $\epsilon_h=1$ and $t \rightarrow 0$.}
\begin{ruledtabular}
\begin{tabular}{lccc}
material & $Q_0$ ($10^6$) & $Q_{TLS,0} / \epsilon_h t $ (1/nm) & $E_s$ (kV/m)\\
\hline

NbTiN & 0.9-1.3 & 330-450 & 0.05\\
Ta & 1.7-3 & 70-90 & 0.1-2 \\
Si (NbTiN) &  & 13-16 & 5\\
Si (Ta) &  & 1.1-1.9 & 2 \\

\end{tabular}
\end{ruledtabular}
\label{table:fitparameters}
\end{table}

The saturation fields of NbTiN are on the order of 50 V/m, see Table
\ref{table:fitparameters}, similar to values for Re and Al
\cite{wang}. For Ta we find a large spread in the saturation fields.
The dielectric loss of NbTiN is clearly smaller than that of Ta. The
substrate surface values are consistent with SiO$_x$. The saturation
field is $E_s \sim 2-5$ kV/m; comparable to measurements on vitreous
silica: $p \sim 1$ D, $T_1 \sim 0.01 - 1$ $\mu$s and $T_2 \sim 1$ ns
\cite{schickfus}. Moreover, a value of $E_s \sim 2-3$ kV/m has been
reported for SiO$_2$ also by Martinis \emph{et al.} \cite{martinis}.
The quality factor of the Si surface layer, assuming $t=3$ nm and
$\epsilon_h=4$, lies around 15-200, which is on the order of the
value of $\sim 200$ reported for SiO$_2$ \cite{martinis}. At high
intensity the quality factors are temperature independent,
suggesting loss other than due to TLS. We include an
intensity-independent fitting term $1/Q_0$ to account for this loss.
We suspect that we reach the level of the intrinsic loss of the
superconductor. For Ta, relaxation times saturate for $T/T_c < 0.2$
\cite{barendsPRL}, suggesting that the quasiparticle density becomes
temperature independent. At $T/T_c = 0.2$ we estimate $Q \sim 10^6$
based on Mattis-Bardeen, on the order of values found for $Q_0$.

The data in Fig. \ref{fig:qstandard} and the analysis provide a
clear guide towards improving the quality factor. We have shown that
Ta suffers from significant microwave loss due to dipole TLS in its
metal surface, i.e. its native oxide. We believe that the presence
of a native oxide is the reason why resonators made of Nb, Ta, Al,
or deposited on top of SiO$_2$, consistently show low quality
factors in the single photon regime
\cite{wang,oconnell,palacios,macha,lindstrom}. In this respect NbTiN
is different, because the metal atoms are bound to nitrogen.
Moreover, resonators with $S=3$ $\mu$m and $W=2$ $\mu$m have quality
factors around $250\cdot 10^3$, nearly a doubling compared to Re on
Si resonators which have quality factors around $150\cdot 10^3$
(loaded, with $Q_c>Q_i$) and are wider ($S=5$ $\mu$m and $W=2$
$\mu$m) \cite{wang}. Nevertheless, the NbTiN resonator quality
factor is significantly limited by the exposed and oxidized Si
surface. Therefore NbTiN has a clean surface compared to Si, as the
metal surface influences the loss much stronger than the exposed
substrate (Fig. \ref{fig:cpw}c). Consequently, removal of the Si
from the gaps will increase the quality factor.

\begin{figure}[!t]
    \centering
    \includegraphics[width=1\linewidth]{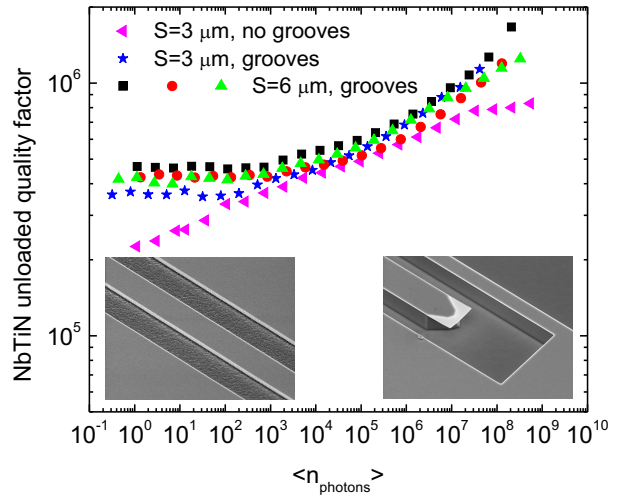}
    \caption{(Color online) The unloaded quality factor versus applied photon number of NbTiN quarterwave
    resonators with the standard geometry and $S=3$ $\mu$m and $W=2$ $\mu$m ($\blacktriangleleft$) (6.2 GHz), and with grooves etched in the exposed Si
    substrate with $S=3$ $\mu$m and $W=2$ $\mu$m ($\bigstar$) (4.7 GHz) and
    $S=6$ $\mu$m and $W=2$ $\mu$m at frequencies of 4.2 ($\blacksquare$), 4.4 (\ding{108}) and 5.2 GHz ($\blacktriangle$). Bath temperature is 60 mK.
    The left inset is a scanning electron microscope image from the standard coplanar waveguide design,
    the right inset shows the etched grooves near the open end of the resonator (the central line width is $S=3$ $\mu$m in both images).
    The cross section of the etched resonators is outlined in Fig. \ref{fig:cpw}a (solid black lines).}
    \label{fig:qgrooves}
\end{figure}

With NbTiN shown to be a superior superconductor, we have redesigned
our resonators to have fewer dielectrics. We have made 50 nm thick
NbTiN resonators, fully straight, which are aligned along the
$\left<110\right>$ axis of the $\left<100\right>$-oriented
HF-cleaned Si wafer. Using KOH wet etching, grooves of 0.9 $\mu$m
deep are etched in the gaps along the full length of the resonators,
see the inset of Fig. \ref{fig:qgrooves}; this removes the substrate
surface from the region with the highest electric field density
(black lines in Fig. \ref{fig:cpw}a).

The NbTiN resonators with grooves etched in the gaps have
significantly higher quality factors, see Fig. \ref{fig:qgrooves}.
In the single photon regime, the quality factor has improved from a
value of $250\cdot 10^3$ for the standard design to an
intensity-independent plateau value of $350\cdot 10^3$ for the
etched resonators, for $S=3$ $\mu$m and $W=2$ $\mu$m. Moreover, this
increase is a clear indication that the Si surface was the limiting
factor also for another reason: the decrease of dielectric has lead
to a decrease in the capacitance $C$ in Eq. \ref{eq:qtls}.
Therefore, if the metal surfaces would dominate the losses, the
quality factors would \emph{decrease}. The intensity-independent
plateau points towards a single surface dominating loss. With the Si
removed, the loss at the single photon level is dominated by the
metal surfaces. Determining which surface is complicated by the
similarity in dependence on intensity and width (Fig.
\ref{fig:cpw}). In the many-photon regime the loss is more due to
the exposed substrate surface, indicated by the higher quality
factors for etched resonators and the high saturation field values.
Finally, when increasing the width to $S=6$ $\mu$m and $W=2$ $\mu$m,
the quality factor improves to around $450\cdot 10^3$. This 30~\%
increase is consistent with our calculation (Fig. \ref{fig:cpw}b)
and shows that further increases can be obtained by widening the
resonator.

With quality factors as high as $470\cdot 10^3$, we estimate single
photon lifetimes of 18 $\mu$s at 4.2 GHz, one order of magnitude
longer than decoherence times measured for superconducting qubits
\cite{bertet,houck}. These long lifetimes make superconducting
resonators, as shown in Fig. \ref{fig:qgrooves}, appealing building
blocks for a quantum processor, as they can be used as quantum
memory elements \cite{leek} and for a quantum bus for long-range
qubit-qubit coupling \cite{sillanpaa,majer}.

To conclude, we have found NbTiN resonators to have a higher quality
factor in the single photon regime than any of the previously
studied superconductors, indicating it has a minimal lossy
dielectric layer. The losses arise largely due to a surface
distribution of two-level systems on the exposed Si substrate. By
removing the substrate from the region with highest electric fields
the quality factor is increased further, showing that using NbTiN
resonators and removing dielectrics is a straightforward route to
high quality factors in the single photon regime.

\begin{acknowledgments}
The authors thank J. M. Martinis and P. Forn-D\'{\i}az for
stimulating discussions. The work was supported by the Pieter
Langerhuizen Lambertuszoon funds of the Royal Holland Society of
Sciences and Humanities and by the EU NanoSciERA project
``Nanofridge''.
\end{acknowledgments}


\begin{thebibliography}{99}


\bibitem{wallraff} A. Wallraff, D. I. Schuster, A. Blais, L.
    Frunzio, R. S. Huang, J. Majer, S. Kumar, S. M. Girvin,
    and R. J. Schoelkopf, Nature \textbf{431}, 162 (2004).

\bibitem{wang} H. Wang, M. Hofheinz, J. Wenner, M. Ansmann, R. C.
    Bialczak, M. Lenander, E. Lucero, M. Neeley, A. D. O'Connell, D.
    Sank, M. Weides, A. N. Cleland, and J. M. Martinis,
    Appl. Phys. Lett. \textbf{95}, 233508 (2009).

\bibitem{oconnell} A. D. O'Connell, M. Ansmann, R. C. Bialczak, M.
    Hofheinz, N. Katz, E. Lucero, C. McKenney, M. Neeley, H. Wang,
    E. M. Weig, A. N. Cleland, and J. M. Martinis,
    Appl. Phys. Lett. \textbf{92}, 112903 (2008).


\bibitem{macha} P. Macha, S. H. W. van der Ploeg, G. Oelsner, E.
    Il'ichev, H.-G. Meyer, S. W\"{u}nsch, and M. Siegel,
    Appl. Phys. Lett. \textbf{96}, 062503 (2010).

\bibitem{lindstrom} T. Lindstr\"{o}m, J. E. Healey, M. S. Colclough,
    C. M. Muirhead, and A. Ya. Tzalenchuk, Phys. Rev. B
    \textbf{80}, 132501 (2009).

\bibitem{palacios} A. Palacios-Laloy, F. Nguyen, F. Mallet, P.
    Bertet, D. Vion, and D. Esteve, J. Low Temp. Phys. \textbf{151},
    1034 (2008).

\bibitem{day} P. K. Day, H. G. LeDuc, B. A. Mazin, A. Vayonakis, and
    J. Zmuidzinas, Nature \textbf{425}, 817 (2003).

\bibitem{barendsapl} R. Barends, H. L. Hortensius, T. Zijlstra, J.
    J.
    A. Baselmans, S. J. C. Yates, J. R. Gao, and T. M. Klapwijk, Appl. Phys.
    Lett. \textbf{92}, 223502 (2008).

\bibitem{barendsthesis} R. Barends, Ph. D. thesis,
    Delft University of Technology, 2009.

\bibitem{numberofphotons} Applied number of photons:
    $\left<n_{photons}\right>=C V_r^2 l /2hf$, with $V_r=2\sqrt{P_{int}Z}$ the
    standing wave voltage, $P_{int}$ the internal resonator power
    \cite{barendsthesis} and $Z$ the waveguide impedance.

\bibitem{phillips} W. A. Phillips, Rep. Prog. Phys. \textbf{50},
    1657 (1987).


\bibitem{martinis} J. M. Martinis, K. B. Cooper, R. McDermott,
    M. Steffen, M. Ansmann, K. D. Osborn, K. Cicak,
    S. Oh, D. P. Pappas, R. W. Simmonds, and C. C. Yu, Phys. Rev. Lett.
    \textbf{95}, 210503 (2005).

\bibitem{gao} J. Gao, M. Daal, A. Vayonakis, S. Kumar, J.
    Zmuidzinas, B. Sadoulet, B. A. Mazin, P. K. Day, and H. G. LeDuc,
    Appl. Phys. Lett. \textbf{92}, 152505 (2008).

\bibitem{finiteelements} P. P. Silvester and R. L. Ferrari,
    \emph{Finite
    elements for electrical engineers}, 2$^{\mathrm{nd}}$ edition (Cambridge
    University Press, 1989).

\bibitem{schickfus} M. von Schickfus and S. Hunklinger, Phys. Lett. \textbf{64A}, 144 (1977).

\bibitem{barendsPRL} R. Barends, J. J. A. Baselmans, S. J. C.
    Yates, J. R. Gao, J. N. Hovenier, and T. M. Klapwijk,
    Phys. Rev. Lett.
    \textbf{100}, 257002 (2008).

\bibitem{bertet} P. Bertet, I. Chiorescu, G. Burkard, K.
    Semba, C. J. P. M. Harmans, D. P. DiVincenzo, and J. E. Mooij,
    Phys. Rev. Lett. \textbf{95}, 257002 (2005).

\bibitem{houck} A. A. Houck, J. A. Schreier, B. R. Johnson, J.
    M. Chow, J. Koch, J. M. Gambetta, D. I. Schuster, L.
    Frunzio, M. H. Devoret, S. M. Girvin, and R. J. Schoelkopf,
    Phys. Rev. Lett. \textbf{101}, 080502 (2008).

\bibitem{leek} P. J. Leek, M. Baur, J. M. Fink, R. Bianchetti, L.
    Steffen, S. Filipp, and A. Wallraff, Phys. Rev. Lett. \textbf{104}, 100504 (2010).

\bibitem{sillanpaa} M. A. Sillanp\"{a}\"{a}, J. I. Park, and R. W.
    Simmonds, Nature \textbf{449}, 438 (2007).

\bibitem{majer} J. Majer, J. M. Chow, J. M. Gambetta, J.
    Koch, B. R. Johnson, J. A. Schreier, L. Frunzio, D. I.
    Schuster, A. A. Houck, A. Wallraff, A. Blais, M. H.
    Devoret, S. M. Girvin, and R. J. Schoelkopf, Nature
    \textbf{449},
    443 (2007).

\end{thebibliography}
\end{document}